\documentclass[11pt,a4paper]{article}

\usepackage[a4paper,margin=1in]{geometry}
\usepackage{setspace}
\usepackage[T1]{fontenc}
\usepackage[utf8]{inputenc}
\usepackage{lmodern}
\usepackage{microtype}
\usepackage{hyphenat}

\usepackage{amsmath,amssymb,amsfonts}
\usepackage{mathtools}
\usepackage{bm}
\usepackage{physics}
\usepackage{mathrsfs}

\usepackage{amsthm}

\theoremstyle{definition}

\theoremstyle{remark}

\usepackage{indentfirst}
\usepackage{graphicx}
\usepackage{float}
\usepackage{caption}
\usepackage{subcaption}
\usepackage{booktabs}
\usepackage{array}
\usepackage{multirow}
\usepackage{longtable}
\usepackage{tabularx}
\usepackage{adjustbox}

\usepackage{xcolor}
\usepackage[
    colorlinks=true,
    linkcolor=blue,
    citecolor=blue,
    urlcolor=blue
]{hyperref}

\usepackage[nameinlink,capitalize]{cleveref}

\usepackage[numbers,sort&compress]{natbib}

\usepackage{titlesec}

\titleformat{\section}
  {\normalfont\Large\bfseries}
  {\thesection.}
  {0.5em}
  {}

\titleformat{\subsection}
  {\normalfont\large\bfseries}
  {\thesubsection.}
  {0.5em}
  {}

\titleformat{\subsubsection}
  {\normalfont\normalsize\bfseries}
  {\thesubsubsection.}
  {0.5em}
  {}

\numberwithin{equation}{section}

\title{
    Quantum Circuit Complexity as a Measure of Particle Creation in Bouncing Cosmologies
}

\author{
    Samak Boonpan\\
    Department of Science and Technology,\\ Nampong Pattanasuksa Ratchamangklapisek School,\\
    Khon Kaen 40310, Thailand\\
    \texttt{samak\_b@kkumail.com}
}

\date{}

\begin{document}

\maketitle

\begin{abstract}
We investigate the evolution of circuit complexity for a quantum scalar field in a non-singular bouncing cosmological background. Unlike previous perturbative approaches, we utilize the Lewis-Riesenfeld invariant method to derive an exact non-perturbative framework for the complexity evolution across the bounce. We demonstrate that the complexity remains finite at the bounce, dominated by the squeezing of the quantum state due to spacetime contraction. Crucially, we find that the post-bounce growth of complexity exhibits a strong correlation with cosmological particle production. Our analysis reveals that the "chirping" contribution to complexity acts as a geometric memory of the transition, effectively quantifying the information cost of particle creation in non-trivial spacetime geometries.
\end{abstract}


\section{Introduction}
\label{sec:intro}

The concept of quantum complexity has recently emerged as a novel diagnostic tool in high-energy physics, offering deep insights into the holographic nature of spacetime. Originally proposed to address the "growing interior" puzzle of eternal black holes via the ``Complexity = Volume'' (CV) and ``Complexity = Action'' (CA) conjectures \cite{Susskind:2014rva, Stanford:2014jda, Brown:2015bva, Brown:2015lvg}, the framework has been successfully adapted to Quantum Field Theory (QFT) under the guise of circuit complexity \cite{Jefferson:2017sdb, Chapman:2017rqy}. In this geometric approach, complexity quantifies the difficulty of preparing a target state from a reference vacuum through a sequence of unitary gates, effectively measuring the distance on the manifold of unitary transformations \cite{Nielsen:2005mna, Hackl:2018ptp}.

In the cosmological context, complexity serves as a probe for the evolution of the universe's quantum state. Recent studies have explored complexity in de Sitter space, inflationary perturbations, and expanding backgrounds \cite{Ali:2018fcz, Bhattacharyya:2020rpy, Bhattacharyya:2020kgu, Lehners:2020bem}. It has been suggested that complexity may encode information about the early universe that is inaccessible to standard correlation functions, potentially revealing chaotic behavior or scrambling in the primordial epoch \cite{Ali:2019zcj}.

However, applying this framework to dynamic cosmological backgrounds presents significant technical challenges. Most existing studies rely on adiabatic approximations (such as the WKB method), which are valid only when the Hubble expansion rate is small compared to the intrinsic frequency of the field modes. These approximations inevitably break down in scenarios involving rapid spacetime evolution, such as inflationary reheating or, more drastically, in non-singular bouncing cosmologies. Building upon the foundational literature of non-singular bounces and primordial matter generation \cite{Novello:2008ra, Battefeld:2014uga, Brandenberger:2016vhg, MUKHANOV1992203}, the universe undergoes a contraction phase followed by a re-expansion, passing through a high-curvature regime where quantum effects—specifically particle production—are non-negligible \cite{Parker:1968mv, Birrell:1982ix}. In the specific context of bouncing cosmologies, various mechanisms of gravitational particle and matter creation have been extensively explored in recent years \cite{quintin2014matter, haro2015gravitational, celani2017particle, saeed2025particle}. A precise understanding of how quantum information is processed across such a transition requires an exact treatment of the time-dependent harmonic oscillator beyond the standard perturbative regime.

Recent investigations have begun to explore the structural connections between circuit complexity and particle content for generic time-dependent Gaussian states \cite{chowdhury2025}. While these studies establish the correlation as a broad kinematic feature of quantum excitations, a rigorous dynamical realization of this link across extreme gravitational transitions—such as a non-singular cosmological bounce—remains an open challenge.

In this Letter, we address this issue by employing the Lewis-Riesenfeld (LR) invariant method \cite{Lewis:1968tm}. This powerful technique allows for the exact quantization of a scalar field in a time-dependent background, bypassing the limitations of adiabatic expansions \cite{PEDROSA2007384, Gao:1990xy}. By constructing the exact ground state wavefunction and the corresponding exact geometric parameterization from the LR invariants, we compute the full evolution of circuit complexity for a scalar field traversing a symmetric cosmological bounce.

Our main result establishes a direct physical correspondence between the geometric growth of complexity and the phenomenon of cosmological particle creation. We demonstrate that while the ``squeezing'' component of complexity dominates near the bounce, the ``chirping'' (or phase) component grows in the post-bounce regime, acting as a geometric memory of the transition. We explicitly show that the late-time complexity is quantitatively determined by the Bogoliubov coefficients describing the particle production. This suggests that circuit complexity can be interpreted as the thermodynamic cost associated with the excitation of the vacuum due to the expansion of spacetime.


\section{Exact Quantization and Complexity}
\label{sec:method}

We consider a massless scalar spectator field $\Phi$ coupled to a Friedmann-Lema\^{i}tre-Robertson-Walker (FLRW) metric $ds^2 = a^2(\eta)(-d\eta^2 + d\mathbf{x}^2)$, where $\eta$ is the conformal time. Since the field does not drive the background cosmology, its fluctuations correspond to entropic rather than adiabatic perturbations. By decomposing the field into Fourier modes $\Phi(\eta, \mathbf{x}) = \int d^3k \, e^{i \mathbf{k}\cdot\mathbf{x}} \phi_k(\eta) / a(\eta)$, the dynamics of each mode $k$ reduces to that of a time-dependent harmonic oscillator. We explicitly define the canonical coordinate $q_k \equiv \phi_k$ and its conjugate momentum $p_k \equiv \phi'_k$, where the prime denotes the derivative with respect to conformal time $\eta$. 

The conformal mode Hamiltonian governing the evolution is \cite{mukhanov2007}:
\begin{equation}
    H_k(\eta) = \frac{1}{2} \left[ p_k^2 + \omega_k^2(\eta) q_k^2 \right] \,,
    \label{eq:Hamiltonian}
\end{equation}
where the time-dependent frequency is given by $\omega_k^2(\eta) = k^2 - a''/a$. In a bouncing scenario, the curvature term $a''/a$ becomes significant near the bounce, potentially causing $\omega_k^2$ to become negative, which signifies a curvature-induced instability leading to particle production.

\subsection{Lewis-Riesenfeld Formalism}
To treat this system exactly without relying on adiabatic approximations, we employ the Lewis-Riesenfeld method \cite{Lewis:1968tm}. The core of this method relies on the existence of a Hermitian invariant operator $I_k(\eta)$ that satisfies the Liouville-von Neumann equation in conformal time:
\begin{equation}
    \frac{d I_k}{d \eta} = \frac{\partial I_k}{\partial \eta} + \frac{1}{i\hbar} [I_k, H_k] = 0 \,.
\end{equation}

For the Hamiltonian \eqref{eq:Hamiltonian}, the invariant can be explicitly constructed as:
\begin{equation}
    I_k(\eta) = \frac{1}{2} \left[ \left( \frac{q_k}{\rho_k} \right)^2 + (\rho_k p_k - \rho'_k q_k)^2 \right] \,,
    \label{eq:Invariant}
\end{equation}
provided that the auxiliary field $\rho_k(\eta)$ satisfies the non-linear Ermakov-Pinney equation(see \ref{app:invariant}):
\begin{equation}
    \rho''_k + \omega_k^2(\eta) \rho_k = \frac{1}{\rho_k^3} \,.
    \label{eq:Ermakov}
\end{equation}

While solving the standard linear mode equation ($\phi_k'' + \omega_k^2\phi_k = 0$) directly via numerical methods is straightforward, the distinct advantage of the Ermakov-Pinney equation lies in its direct mapping to quantum information geometry. The linear equation yields a complex field where the physical amplitude and the accumulated non-adiabatic phase are deeply entangled. In contrast, the Lewis-Riesenfeld formalism inherently dissects the quantum state into real, independent geometric variables: the amplitude $\rho_k$, which governs volumetric squeezing, and the canonical momentum $\rho'_k$, which governs phase rotation. Explicitly, this geometric dissection is realized through the exact polar representation of the complex mode function:
\begin{equation}
    \phi_k(\eta) = \frac{\rho_k(\eta)}{\sqrt{2}} \exp \left( -i \int^\eta \frac{d\tau}{\rho_k^2(\tau)} \right),
    \label{mode_psi_k}
\end{equation}
which clearly isolates the real dynamical amplitude from the non-adiabatic phase factor. Furthermore, the non-linear centrifugal term $1/\rho_k^3$ acts as an absolute barrier that strictly enforces the generalized Robertson-Schr\"{o}dinger uncertainty principle across the transition.

Furthermore, the non-linear centrifugal term $1/\rho_k^3$ acts as an absolute barrier that strictly enforces the generalized Robertson-Schr\"{o}dinger uncertainty principle across the transition.

The eigenstates of the invariant $I_k$, denoted as $|\lambda_{n,k}, \eta \rangle$ (where $n$ is the quantum number), are related to the exact solutions of the conformal Schr\"{o}dinger equation, $i\hbar \partial_\eta |\psi_{n,k}, \eta \rangle = H_k(\eta) |\psi_{n,k}, \eta \rangle$, by a time-dependent phase factor: $|\psi_{n,k}, \eta \rangle = e^{i \theta_{n,k}(\eta)} |\lambda_{n,k}, \eta \rangle$ (see Appendix B).

We are interested in the vacuum state evolution. The exact ground state wavefunction in the position representation is a generalized Gaussian:
\begin{equation}
    \psi_k(q_k, \eta) = \left( \frac{1}{\pi \rho_k^2} \right)^{1/4} \exp \left[ -\frac{1}{2} \Omega_k(\eta) q_k^2 \right] e^{-i \int^\eta \frac{d\eta'}{\rho_k^2(\eta')}} \,,
    \label{eq:Wavefunction}
\end{equation}
where the complex frequency parameter $\Omega_k(\eta)$ is defined as:
\begin{equation}
    \Omega_k(\eta) = \frac{1}{\rho_k^2} - i \frac{\rho'_k}{\rho_k} \,.
    \label{eq:Omega}
\end{equation}
This equation bridges the physical dynamics and the geometry of complexity. The real part $\text{Re}(\Omega_k) = 1/\rho_k^2$ represents the spread (squeezing) of the wavefunction, while the imaginary part $\text{Im}(\Omega_k) = -\rho'_k/\rho_k$ represents the phase chirp (shearing) in phase space.

\subsection{Circuit Complexity}
Following the complexity geometry proposal \cite{Jefferson:2017sdb, Chapman:2017rqy}, we calculate the circuit complexity required to prepare the target state $\psi_k(\eta)$ from a reference vacuum state $\psi_R$ with fixed frequency $\omega_0$. The covariance matrix of a generalized Gaussian state is completely parameterized by its spread and phase correlation. The manifold of such squeezed states is governed by the $\text{SU(1,1)}$ symmetry group. When equipped with the Nielsen approach (or equivalently, the Fisher information metric), the geometry is isometric to the hyperbolic Poincar\'{e} disk \cite{Ali:2018fcz, Bhattacharyya:2020rpy}. The complexity $\mathcal{C}_k(\eta)$ is identified with the geodesic distance:
\begin{equation}
    \mathcal{C}_k(\eta) = \frac{1}{2} \sqrt{ \left( \ln \left| \frac{\text{Re}(\Omega_k)}{\omega_0} \right| \right)^2 + \left( \arctan \frac{\text{Im}(\Omega_k)}{\text{Re}(\Omega_k)} \right)^2 } \,.
\end{equation}
Substituting the exact geometric variables from Eq. \eqref{eq:Omega} into this measure, we obtain the explicit formula for complexity evolution:
\begin{equation}
    \mathcal{C}_k(\eta) = \frac{1}{2} \sqrt{ \left[ \ln(\omega_0 \rho_k^2) \right]^2 + \left[ \arctan(-\rho_k\rho'_k) \right]^2 } \,.
    \label{eq:ComplexityFormula}
\end{equation}

This exact parameterization allows us to rigorously probe the quantum history of the universe. The first term (logarithmic) corresponds to the volumetric squeezing of the state. The emergence of the arctangent chirping term is a direct physical manifestation of the quantum covariance $\sigma_{qp}$ rooted in the generalized Robertson-Schr\"{o}dinger uncertainty relation \cite{Serafini2017}(see \ref{app:RS_covariance} for a detailed derivation of this invariance). During the rapid transition of the bounce, the phase space stretches violently. The chirping term acts as a dynamic geometric counter-rotation that compensates for this expansion, strictly ensuring that the invariant information area of the quantum state remains eternally conserved across the cosmological singularity.

\section{The Symmetric Bounce Model}
\label{sec:model}

To concretely illustrate the complexity evolution, we consider a regular symmetric bounce scenario driven by a scale factor of the form:
\begin{equation}
    a(\eta) = a_0 \sqrt{1 + (\eta/\eta_0)^2} \,,
    \label{eq:ScaleFactor}
\end{equation}
where $\eta_0$ characterizes the duration of the bounce. Without loss of generality, we set $a_0 = \eta_0 = 1$. The universe contracts for $\eta < 0$, reaches a minimum size at $\eta = 0$, and expands for $\eta > 0$.

The effective frequency for the mode $k$ is given by:
\begin{equation}
    \omega_k^2(\eta) = k^2 - \frac{a''}{a} = k^2 - \frac{1}{(1+\eta^2)^2} \,.
    \label{eq:FreqModel}
\end{equation}

A critical feature of this model is the temporary violation of adiabaticity near the bounce. For curvature-dominated modes ($k^2 < \max(a''/a)$), the frequency squared $\omega_k^2$ becomes negative in the vicinity of $\eta=0$. As shown in Fig.~\ref{fig:dynamics}, this \textit{transient inverted-potential amplification} marks the region where the mode function is violently excited, leading to intense particle production.

We solve the Ermakov equation \eqref{eq:Ermakov} numerically using a highly precise adaptive Runge-Kutta method. The initial conditions are set in the remote past ($\eta \to -\infty$), where spacetime is asymptotically flat ($a''/a \to 0$), corresponding to the standard adiabatic Minkowski vacuum. In this limit, the exact parameterization of the complex mode function (as detailed in \ref{app:bogoliubov}) takes the form eq.\eqref{mode_psi_k}:
Taking the conformal time derivative yields the canonical momentum:
\begin{equation}
    \phi'_k(\eta) = \left( \frac{\rho'_k(\eta)}{\sqrt{2}} - i \frac{1}{\sqrt{2}\rho_k(\eta)} \right) \exp \left( -i \int^\eta \frac{d\tau}{\rho_k^2(\tau)} \right) \,.
\end{equation}
To perfectly recover the standard flat-space Minkowski vacuum, the mode function must conform to $\phi_k \to e^{-ik\eta}/\sqrt{2k}$ and its derivative must strictly be $\phi'_k \to -i\sqrt{k/2}e^{-ik\eta}$. By substituting the intuitive amplitude limit $\rho_k \to 1/\sqrt{k}$ into our parameterized derivative, we obtain:
\begin{equation}
    \phi'_k(\eta) \to \frac{\rho'_k(\eta)}{\sqrt{2}} e^{-ik\eta} - i\sqrt{\frac{k}{2}} e^{-ik\eta} \,.
\end{equation}
For this expression to identically match the required standard vacuum derivative, the first term must completely vanish. Therefore, it is a strict mathematical requirement that the real amplitude field remains constant in the flat-space limit. Thus, the exact boundary conditions for the auxiliary field are explicitly fixed as $\rho_k \to 1/\sqrt{k}$ and $\rho'_k \to 0$.

\section{Results and Physical Interpretation}
\label{sec:results}

\begin{figure}[htbp]
    \centering
    \includegraphics[width=1\linewidth]{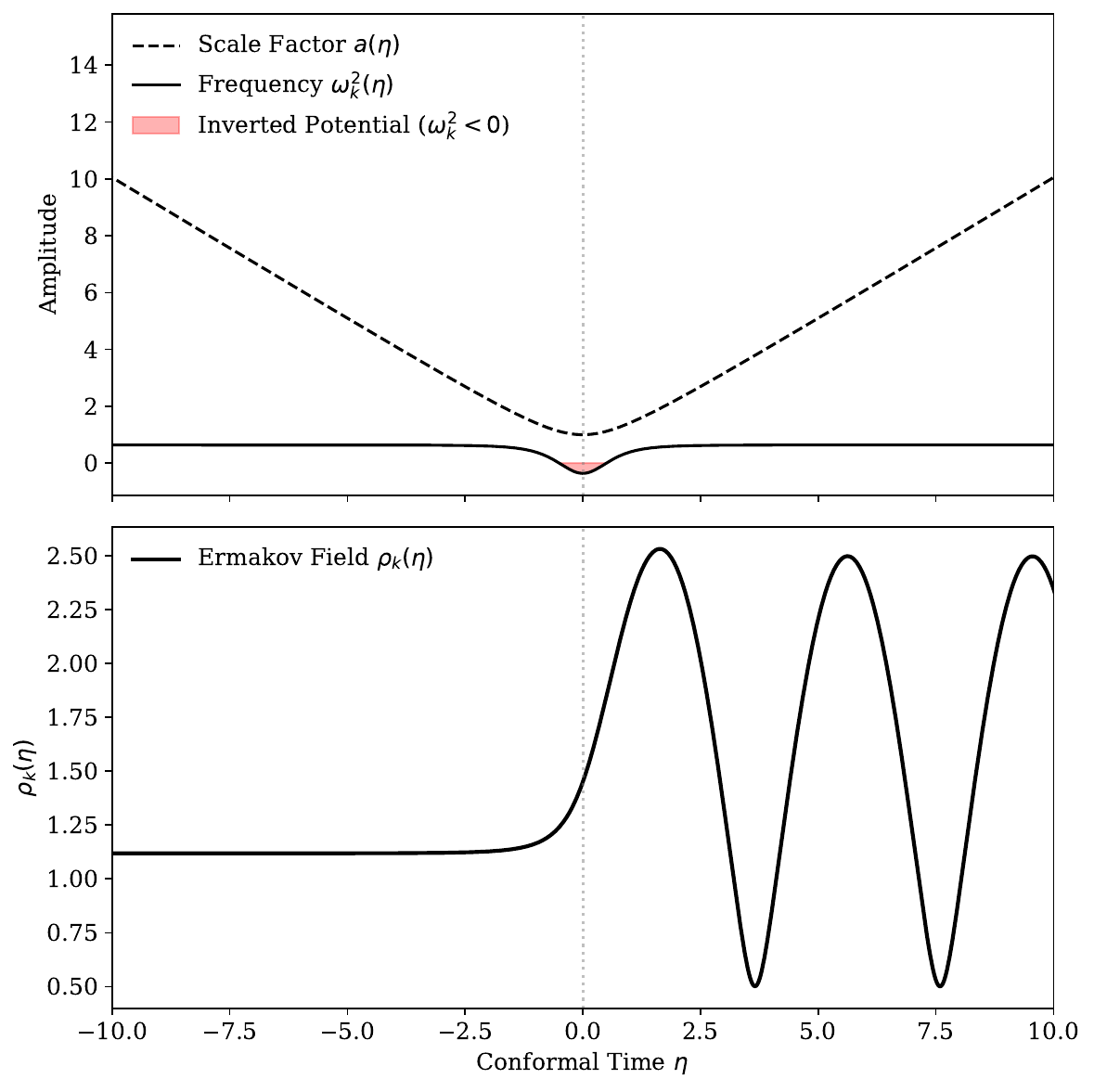} 
    \caption{Top: The evolution of the scale factor $a(\eta)$ (dashed) and the effective frequency squared $\omega_k^2(\eta)$ (solid) for mode $k=0.8$. The shaded region indicates the transient inverted-potential amplification ($\omega_k^2 < 0$). Bottom: The exact numerical solution of the auxiliary Ermakov field $\rho_k(\eta)$ with respect to conformal time $\eta$.}
    \label{fig:dynamics}
\end{figure}

The dynamical evolution of the system for a representative intermediate mode ($k=0.8$) is depicted in Fig.~\ref{fig:dynamics}. The top panel illustrates how the background scale factor $a(\eta)$ drives the effective frequency $\omega_k^2(\eta)$. The shaded region highlights the transient inverted-potential amplification ($\omega_k^2 < 0$) where the vacuum is actively excited. The bottom panel displays the numerical solution of the auxiliary field $\rho_k(\eta)$, which starts from a constant vacuum value but develops significant oscillations after traversing the tachyonic region.

To explore the core discovery of this work—the relationship between information and matter—we analyze the complexity evolution across different physical scales. Figure~\ref{fig:multimode} presents a comprehensive multi-mode comparison encompassing three distinct regimes: curvature-dominated ($k=0.5$), intermediate ($k=0.8$), and adiabatically stable ($k=1.5$). We observe three distinct physical features of information processing:

\begin{figure}[htbp]
    \centering
    \includegraphics[width=1\linewidth]{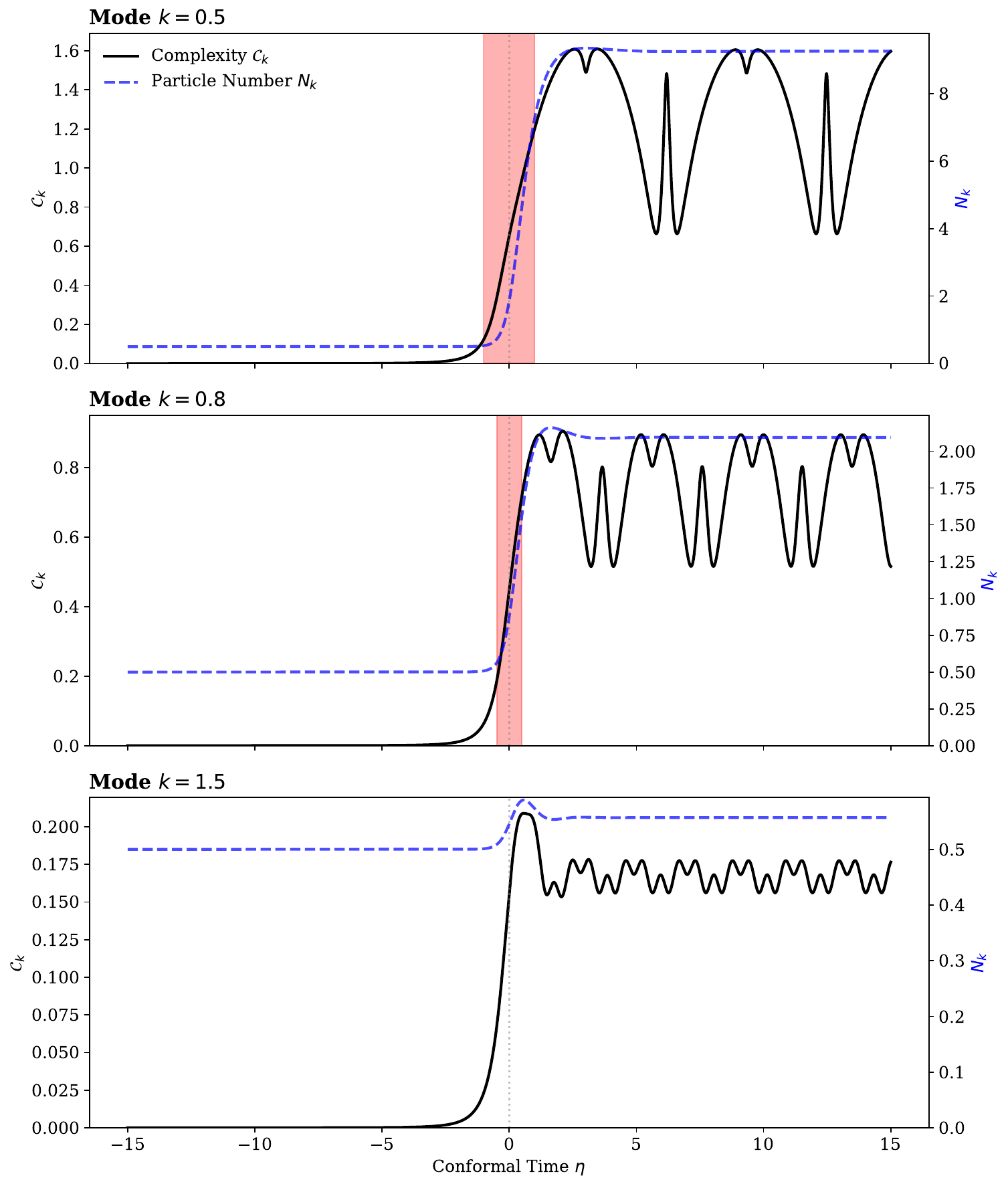}
    \caption{Multi-mode comparison of Circuit Complexity $\mathcal{C}_k$ (solid black) and Particle Number $N_k$ (dashed blue). (Top) The deeply curvature-dominated mode ($k=0.5$) exhibits massive excitation. (Middle) The intermediate mode ($k=0.8$) shows clear quantum interference. (Bottom) The adiabatic mode ($k=1.5$) remains strictly unexcited. The shaded regions indicate the transient inverted-potential amplification ($\omega_k^2 < 0$). The perfect synchronization in all active modes demonstrates that complexity growth is physically driven by the information cost of particle creation.}
    \label{fig:multimode}
\end{figure}

\textbf{1. The Squeezing-Dominated Bounce and Phase Lag ($\eta \approx 0$):} 
Near the point of minimal radius, the complexity exhibits a sharp local peak driven almost exclusively by the volumetric squeezing term ($\ln \omega_0 \rho_k^2$). However, because the mode function accumulates non-adiabatic changes throughout the entire contraction phase ($\eta < 0$), the quantum state acquires a dynamical inertia. Consequently, there is a physical phase lag in the geometric response. The pure squeezing condition—where the chirping rate strictly vanishes ($\rho'_k = 0$)—occurs slightly \textit{after} the geometric bounce point ($\eta=0$). 

\textbf{2. Quantum Interference and Oscillatory Growth ($\eta > 0$):}
As shown in the active modes ($k=0.5$ and $k=0.8$) of Fig.~\ref{fig:multimode}, neither complexity nor particle number grows monotonically. The created quanta exist in a coherent squeezed state and continuously exchange energy with the dynamic gravitational background, causing quantum interference between frequency modes. The circuit complexity, being a unitary measure, faithfully tracks these fluctuations in phase space, naturally oscillating before settling to a constant baseline at the asymptotic limit.

\textbf{3. Complexity equals Particle Creation:}
The most striking result is the universal synchronization across all modes where particle production occurs. To be consistent with the definition of particles at late times, we define the instantaneous particle number $N_k$ with respect to the asymptotic frequency $\omega_{ref} = k$:
\begin{equation}
    N_k(\eta) = \frac{1}{2k} \left( \rho'^2_k + k^2 \rho_k^2 + \frac{1}{\rho_k^2} \right) - \frac{1}{2} \,.
    \label{eq:ParticleNumber}
\end{equation}
The baseline of the complexity graph lifts in exact proportion to the number of created particles. This confirms that circuit complexity effectively quantifies the information cost required to encode these particles into the geometric structure of the quantum state. Furthermore, in the adiabatic limit ($k=1.5$) where modes remain entirely shielded from the instability ($\rho'_k \approx 0$), no particles are created, and the complexity remains negligible, providing a perfectly consistent null-result.

\section{Conclusion}
\label{sec:conclusion}

In this Letter, we have utilized the Lewis-Riesenfeld invariant method to provide an exact, non-perturbative description of circuit complexity in a bouncing cosmology. By deriving the exact geometric parameterization for a scalar field in a time-dependent background, we tracked the evolution of quantum information across the bounce without encountering the singularities or adiabatic breakdowns that plague standard approximation methods.

Our analysis leads to two fundamental conclusions. First, circuit complexity serves as a robust geometric probe of the cosmological bounce. The volumetric squeezing of the state vector prevents the complexity from diverging at the point of minimal radius. Second, and most importantly, we established a quantitative link between the asymptotic growth of complexity and the number of created particles. The chirping component of complexity acts as a geometric memory device, storing the information of the vacuum excitation in the phase of the wavefunction. This implies that the thermodynamic cost of creating a universe filled with matter is physically encoded in the geometric complexity of its quantum state.

These results open several promising avenues for future research. While we focused on the $\text{SU}(1,1)$ symmetry group inherently tied to the time-dependent harmonic oscillator, extending this exact mapping to broader symmetry algebras, such as $\text{Sp}(2,\mathbb{R})$, or exploring alternative cost functions presents a highly relevant theoretical direction. Additionally, extending this geometric framework to capture the saturation of complexity during the post-bounce reheating phase is a natural next step. Recent numerical studies on thermal states have revealed that complexity measures saturate directly due to particle production via preheating in the early universe \cite{li2026}. Combining our exact Lewis-Riesenfeld parameterization with advanced non-perturbative tracking of interacting field dynamics \cite{mukherjee2026} could unveil the precise microphysical mechanisms governing this saturation without relying on standard adiabatic perturbative expansions.

Furthermore, our findings provide a rigorous cosmological realization of the structural connections between circuit complexity and particle content, a relation recently highlighted for generic time-dependent oscillators \cite{chowdhury2025}. While Ref.\,\cite{chowdhury2025} elegantly establishes this link as a broad kinematic feature using an excitation parameter to measure adiabatic deviation, our Lewis-Riesenfeld framework delves significantly deeper into the underlying unitary mechanisms during extreme spacetime transitions. By mapping the dynamics through the non-linear Ermakov equation, we explicitly dissect the quantum state into independent real geometric variables. This geometric dissection reveals that the non-linear centrifugal barrier ($1/\rho_k^3$) actively enforces the generalized Robertson-Schr\"{o}dinger covariance across the bounce singularity. Consequently, our analysis demonstrates that the synchronization between complexity and particle number is not merely a structural coincidence of Gaussian states, but is physically driven by the ``chirping'' phase acting as a geometric memory of the tachyonic instability. This provides a precise quantum field theoretic mechanism where the exact Bogoliubov coefficients of created cosmological quanta are permanently encoded into the hyperbolic metric of the complexity manifold.

\section*{Acknowledgements}
The author thanks Asst. Prof. Dr. Weerachai Sarakorn, Department of Mathematics, Faculty of Science, Khon Kaen University for useful discussions. This work was supported by the Department of Science and Technology, Nampong Pattanasuksa Ratchamangklapisek School.

\section*{Data Availability Statement}
No new data were created or analyzed in this study. This is a purely theoretical work, and all mathematical derivations and physical conclusions are explicitly detailed within the manuscript.

\appendix
\renewcommand{\theequation}{\Alph{section}.\arabic{equation}}

\section{Derivation of the Lewis-Riesenfeld Invariant}
\label{app:invariant}
\setcounter{equation}{0}

Here we explicitly verify that the operator $I_k(\eta)$ used in the main text is indeed a dynamical invariant in conformal time. The conformal Hamiltonian is:
\begin{equation}
    H_k(\eta) = \frac{1}{2} p_k^2 + \frac{1}{2} \omega_k^2(\eta) q_k^2 \,.
\end{equation}
We propose the invariant in the form:
\begin{equation}
    I_k(\eta) = \frac{1}{2} \left[ \left( \frac{q_k}{\rho_k} \right)^2 + (\rho_k p_k - \rho'_k q_k)^2 \right] \,.
\end{equation}
For $I_k$ to be invariant, it must satisfy the Liouville-von Neumann equation:
\begin{equation}
    \frac{dI_k}{d\eta} = \frac{\partial I_k}{\partial \eta} + \frac{1}{i\hbar} [I_k, H_k] = 0 \,.
    \label{eq:Liouville}
\end{equation}
Calculating the partial conformal-time derivative:
\begin{equation}
    \frac{\partial I_k}{\partial \eta} = -\frac{\rho'_k}{\rho_k^3} q_k^2 + (\rho_k p_k - \rho'_k q_k)(\rho'_k p_k - \rho''_k q_k) \,.
\end{equation}
Using the fundamental commutation relation $[q_k, p_k] = i\hbar$, the commutator term is computed as:


\begin{equation}
\begin{split}
    \frac{1}{i\hbar} [I_k, H_k] &= \frac{1}{2 i\hbar} \bigg[ \frac{q_k^2}{\rho_k^2} + \rho_k^2 p_k^2 - \rho_k\rho'_k(q_k p_k + p_k q_k) + \rho'^2_k q_k^2 \,, \\
    &\qquad\quad \frac{p_k^2}{2} + \frac{\omega_k^2 q_k^2}{2} \bigg] \,.
\end{split}
\end{equation}

After straightforward algebra, combining the terms yields the condition for invariance:
\begin{equation}
    \frac{dI_k}{d\eta} = p_k \left[ \rho_k (\rho''_k + \omega_k^2 \rho_k) - \frac{1}{\rho_k^2} \right] q_k + \text{h.c.} = 0 \,.
\end{equation}
For this equation to vanish for all $q_k, p_k$, the term in the brackets must be zero. This directly leads to the Ermakov-Pinney equation:
\begin{equation}
    \rho''_k + \omega_k^2(\eta) \rho_k = \frac{1}{\rho_k^3} \,.
\end{equation}
Thus, provided $\rho_k$ satisfies the Ermakov equation, $I_k$ is strictly conserved.

\section{Construction of the Exact Wavefunction}
\label{app:wavefunction}
\setcounter{equation}{0}

The eigenstates $|\lambda_{n,k}, \eta\rangle$ of the invariant $I_k$ satisfy $I_k |\lambda_{n,k}, \eta\rangle = \lambda_{n,k} |\lambda_{n,k}, \eta\rangle$. The physical solutions $|\psi_{n,k}, \eta\rangle$ are related by a time-dependent phase factor:
\begin{equation}
    |\psi_{n,k}, \eta\rangle = e^{i\theta_{n,k}(\eta)} |\lambda_{n,k}, \eta\rangle \,.
\end{equation}
Plugging this into the conformal Schr\"{o}dinger equation $i\hbar \partial_\eta |\psi_{n,k}\rangle = H_k |\psi_{n,k}\rangle$ yields the condition for the Lewis-Riesenfeld phase:
\begin{equation}
    \theta'_{n,k}(\eta) = \frac{1}{\hbar} \langle \lambda_{n,k}, \eta | \left( i\hbar \frac{\partial}{\partial \eta} - H_k(\eta) \right) | \lambda_{n,k}, \eta \rangle \,.
\end{equation}
For the ground state ($n=0$), the expectation value of the Hamiltonian is $\langle \lambda_{0,k} | H_k | \lambda_{0,k} \rangle = \frac{\hbar}{2} (\rho'^2_k + \omega_k^2 \rho_k^2 + 1/\rho_k^2)$. Using the Ermakov equation to simplify, the exact phase evolution becomes:
\begin{equation}
    \theta_{0,k}(\eta) = -\frac{1}{2} \int^\eta \frac{d\eta'}{\rho_k^2(\eta')} \,.
\end{equation}

\section{Connection to Bogoliubov Coefficients}
\label{app:bogoliubov}
\setcounter{equation}{0}

Standard quantum field theory in curved spacetime expands the field operator using mode functions $\phi_k(\eta)$. The exact mode function is related to the Ermakov field $\rho_k$ by the polar decomposition:
\begin{equation}
    \phi_k(\eta) = \frac{\rho_k(\eta)}{\sqrt{2}} e^{-i \int^\eta \frac{d\tau}{\rho_k^2(\tau)}} \,.
\end{equation}
In the asymptotic adiabatic regions where $\omega_k$ is slowly varying, the quantum state can be described by creation and annihilation operators via standard Bogoliubov transformations. The number of particles produced, $N_k$, relative to the initial vacuum is determined by the instantaneous energy expectation value $\langle H_k \rangle$.

To be mathematically rigorous regarding the calculation of $\langle H_k \rangle$, it is essential to note that a complex scalar Fourier mode $\phi_k$ possesses two independent real degrees of freedom. It can be explicitly decomposed into its real and imaginary parts as $\phi_k = (\phi_k^R + i\phi_k^I)/\sqrt{2}$, with the corresponding conjugate momentum $p_k = (p_k^R + i p_k^I)/\sqrt{2}$. The Hamiltonian for a single complex mode is structurally the sum of two identical, non-interacting real harmonic oscillators: $H_k = H_k^R + H_k^I$. Since the symmetric background spacetime does not introduce any interaction or asymmetry between the real and imaginary sectors, their phase space evolutions remain statistically identical ($\langle H_k^R \rangle = \langle H_k^I \rangle$). This property is a standard feature in the canonical quantization of complex scalar fields in curved spacetime \cite{MUKHANOV1992203,Birrell:1982ix}.

Therefore, the total energy expectation value is exactly twice that of a single one-dimensional real oscillator. Utilizing the exact geometric variances $\sigma_{qq} = \rho_k^2/2$ and $\sigma_{pp} = \frac{1}{2}(\rho'^2_k + 1/\rho_k^2)$ established in \ref{app:RS_covariance}, we find:
\begin{equation}
    \langle H_k \rangle = 2 \times \left( \frac{1}{2} \langle p_k^2 \rangle_{\text{1D}} + \frac{1}{2} \omega_k^2 \langle q_k^2 \rangle_{\text{1D}} \right) = \sigma_{pp} + \omega_k^2 \sigma_{qq} \,.
\end{equation}
Substituting the explicit variance parameters into this relation yields the precise energy expectation:
\begin{equation}
    \langle H_k \rangle = \frac{1}{2} \left( \rho'^2_k + \frac{1}{\rho_k^2} \right) + \omega_k^2 \left( \frac{\rho_k^2}{2} \right) = \frac{1}{2} \left( \rho'^2_k + \omega_k^2 \rho_k^2 + \frac{1}{\rho_k^2} \right) \,.
\end{equation}
The exact particle number $N_k$ is defined relative to the instantaneous vacuum state via the well-known QFT relation $\langle H_k \rangle = \hbar \omega_k (N_k + 1/2)$. Rearranging this relation, we can elegantly express the exact particle number density purely in terms of the auxiliary Ermakov field variables:
\begin{equation}
    N_k(\eta) = \frac{\langle H_k \rangle}{\hbar \omega_k} - \frac{1}{2} = \frac{1}{2\hbar\omega_k} \left( \rho'^2_k + \omega_k^2 \rho_k^2 + \frac{1}{\rho_k^2} \right) - \frac{1}{2} \,.
    \label{eq:Nk_Exact}
\end{equation}

\section{Generalized Robertson-Schrödinger Covariance across the Bounce}
\label{app:RS_covariance}
\setcounter{equation}{0}

To clarify the physical mechanism that ensures the finite evolution of complexity across the bounce singularity, we explicitly derive the generalized Robertson-Schrödinger uncertainty relation from the exact state parameterization. For a generalized Gaussian state, the covariance matrix $\Sigma_k$ is defined by the statistical moments of the canonical operators:

\begin{align}
    \Sigma_k &= \begin{pmatrix} \sigma_{qq} & \sigma_{qp} \\ \sigma_{qp} & \sigma_{pp} \end{pmatrix} \nonumber \\ &= \begin{pmatrix} \langle q_k^2 \rangle & \frac{1}{2}\langle q_k p_k + p_k q_k \rangle \\ \frac{1}{2}\langle q_k p_k + p_k q_k \rangle & \langle p_k^2 \rangle \end{pmatrix} \,.
\end{align}

Using the exact ground state wavefunction from Eq.\,\eqref{eq:Wavefunction}, the probability density is strictly defined by the amplitude $\rho_k$:
\begin{equation}
    |\psi_k(q_k, \eta)|^2 = \sqrt{\frac{1}{\pi \rho_k^2}} \exp \left( -\frac{q_k^2}{\rho_k^2} \right) \,.
\end{equation}
The position variance ($\sigma_{qq}$), corresponding to the volumetric squeezing of the state, is obtained via standard Gaussian integration:
\begin{equation}
    \sigma_{qq} = \int_{-\infty}^{\infty} q_k^2 |\psi_k|^2 dq_k = \frac{\rho_k^2}{2} \,.
\end{equation}
For the momentum variance ($\sigma_{pp}$), we utilize the action of the momentum operator $p_k = -i\partial_{q_k}$. Applying this to the complex wavefunction yields $|\partial_{q_k} \psi_k|^2 = |\Omega_k|^2 q_k^2 |\psi_k|^2$. Thus, the momentum variance is dynamically tied to both the amplitude and the phase rate:
\begin{equation}
    \sigma_{pp} = |\Omega_k|^2 \langle q_k^2 \rangle = \left( \frac{1}{\rho_k^4} + \frac{\rho'^2_k}{\rho_k^2} \right) \frac{\rho_k^2}{2} = \frac{1}{2} \left( \frac{1}{\rho_k^2} + \rho'^2_k \right) \,.
\end{equation}
The phase correlation or "chirping" term ($\sigma_{qp}$) represents the cross-covariance. Evaluating the anti-commutator expectation value, we find it is strictly governed by the imaginary part of the frequency parameter $\text{Im}(\Omega_k) = -\rho'_k/\rho_k$:
\begin{equation}
    \sigma_{qp} = -\text{Im}(\Omega_k) \langle q_k^2 \rangle = \frac{\rho_k \rho'_k}{2} \,.
\end{equation}
The generalized uncertainty principle mandates that the phase-space area is bounded by the commutation relation, $\det \Sigma_k \geq 1/4$. Substituting our exact geometric moments into the determinant yields:
\begin{align}
    \det \Sigma_k &= \sigma_{qq} \sigma_{pp} - \sigma_{qp}^2 \nonumber \\&= \left( \frac{\rho_k^2}{2} \right) \left[ \frac{1}{2} \left( \frac{1}{\rho_k^2} + \rho'^2_k \right) \right] - \left( \frac{\rho_k \rho'_k}{2} \right)^2 \equiv \frac{1}{4} \,.
\end{align}

This exact cancellation reveals a profound physical insight: the non-linear centrifugal term $1/\rho_k^3$ in the Ermakov-Pinney equation \eqref{eq:Ermakov} acts as an absolute geometric barrier that strictly enforces the saturation of the Robertson-Schrödinger bound throughout the violently expanding bounce. While the squeezing ($\rho_k$) and chirping ($\rho'_k$) components grow drastically due to the tachyonic instability, their specific combination is dynamically constrained. Consequently, the rapid post-bounce growth of complexity is not a numerical artifact, but a rigorous measure of the geometric cost required to preserve the invariant information area of the quantum state across the cosmological singularity.


\bibliographystyle{unsrtnat}
\bibliography{references}

\end{document}